\documentclass[letterpaper]{JHEP3}
\usepackage{graphicx}
\title{Accelerated expansion and the Goldstone ghost}
\author{B. Holdom\\ Department of Physics\\University of Toronto\\Toronto ON Canada M5S1A7\\Email: \email{bob.holdom@utoronto.ca}}
\abstract{A vacuum instability due to a massless ghost in a hidden sector can lead to an effective equation of state for dark energy that changes smoothly from $w=-3/2$ at large redshifts, to $w\approx-1.2$ today, to $w=-1$ in the future. We discuss how this ghost can be the Goldstone boson of Lorentz symmetry breaking, and we find that this breaking in the hidden sector should occur at a scale below $\sim$10 KeV. The normal particles that are produced along with the ghosts are then predominantly neutrinos.}
\begin{document}
\section{Introduction}
It would be disconcerting if any real undertanding of the current accelerated expansion of the universe was blocked by our profound ignorance of the cosmological constant. This motivates us to try to approach the dark energy problem in a manner quite unrelated to the cosmological constant problem. In particular we will consider a very weak vacuum instability as responsible for the origin of dark energy, which appears late in the evolution of the universe. The instability will arise through the appearance of an effective ghost, a degree of freedom with the wrong-sign kinetic term. The presence of a ghost implies that the vacuum is unstable to production of ghosts plus normal matter while conserving energy. One can view this simply as an out of equilibrium situation, where the system is trying to choose a more suitable configuration.

We would like to consider this apparent runaway behavior in more detail when its effect on cosmic evolution is also considered. We will only keep one degree of freedom of gravity, that of the scale factor $a(t)$ in the spatially flat FRW metric, $ds^2=dt^2-a(t)^2d\mathbf{x}^2$. We will model the production of ghost and nonghost degrees of freedom as increasing densities of two perfect fluids, one ghost-like and one normal.  The respective equations of state are $p_g=w_g \rho_g$ and $p_n=w_n \rho_n$, where $p_g$ and $\rho_g$ are negative and $p_n$ and $\rho_n$ are positive. The normal fluid becomes a component of the total normal matter content of the universe, but we will continue to distinguish it from the rest of the normal matter.

If the ghost fluid was truly decoupled from everything else then it would simply dilute away with the expansion along with the normal matter. It is only when there is interaction causing nonconservation (production) of the ghost fluid can there be more dramatic effects. At the very least it can be expected that gravity itself will induce the interaction, and in particular a virtual graviton exchange can produce a pair of ghost particles and a pair of normal particles \cite{cline}. We will also be led to consider a nongravitational interaction, but in either case we refer to the sector containing the ghost as a hidden sector.

We introduce a nonconservation of the two fluids so that $T_n^{\mu\nu}+T_g^{\mu\nu}$ is conserved,
\begin{equation}
D_\mu T_n^{\mu\nu}=-D_\mu T_g^{\mu\nu}=\varepsilon U^\nu,
\label{e6}\end{equation}
where $U^\nu$ is a velocity 4-vector and $\varepsilon$ is a positive finite constant. $U^\nu$ can be used to define the rest frame of the ghost fluid. But the question is how this preferred frame gets chosen in the first place, given that the underlying process is vacuum $\rightarrow$ something. Since the vacuum of the normal sector is Lorentz invariant, we see already that there must be something intrinsically Lorentz violating about the vacuum state of the hidden sector.

We will assume that $U^\nu$ is aligned with the cosmological rest frame. In this frame we assume a constant ghost particle production rate per unit volume $\Gamma_g$ and an average (negative) energy per ghost particle $\overline{E}_g$. Then $\varepsilon=-\overline{E}_g\Gamma_g$, with a similar relation for the normal particles. Equations (\ref{e6}) along with the Einstein equations yield interesting solutions for $\delta w\equiv w_g-w_n > 0$. The $\delta w < 0$ case leads quickly to curvature singularities, while the $\delta w = 0$ case is not interesting for gravity. There, the solution is simply $a(t)=1$ and $\rho_n=-\rho_g=\rho_0+\varepsilon t$ in the absence of other matter.

\section{Accelerated expansion}

In the more interesting $\delta w\equiv w_g-w_n>0$ case the net pressure of the two fluids is negative. This will only add to the expansionary tendency of the universe. But due to the expansion and the relative equation of states, the energy in the negative energy fluid will dilute more rapidly than the positive energy fluid. The net result is the production of positive energy and negative pressure. This energy will eventually dominate the energy density of the universe, in a time that depends on $\varepsilon$ and $\delta w$. A steady state situation is reached where the production is balanced by the dilution from the expansion. The result is de Sitter expansion, $a(t)\propto\exp(H_\mathrm{ss}t)$, with
\begin{equation}
H_\mathrm{ss}=\left({\frac {\varepsilon\delta w}{ 9( 1+ w_n ) 
 ( 1+ w_g )m_\mathrm{Pl}^2 }}\right)^{1/3},
\end{equation}
\begin{equation}
\rho_n={\varepsilon\over 3(1+w_n)H_\mathrm{ss}},\hspace{4ex}\rho_g=-{\varepsilon\over 3(1+w_g)H_\mathrm{ss}},
\end{equation}
and $\rho_n+\rho_g=3m_\mathrm{Pl}^2H_\mathrm{ss}^2$ (where the reduced Planck mass is $m_\mathrm{Pl}=1/\sqrt{8\pi G}$).  This steady state solution will be the endpoint of a wide range of initial conditions involving other matter content of the universe.

This fixed point solution is a generalization of the old Steady State model \cite{hoyle} which had $w_g=1$ and $w_n=0$. This steady state solution has also been discussed recently in \cite{gibbons} for $w_g=1$ and arbitrary $w_n$. In these discussions $w_g=1$ arises from a free massless scalar ghost field with solution $\dot{\phi_g}=$ constant. As in our case the ghost $T_g^{\mu\nu}$ is not conserved, but in the old picture the actual source of the nonconservation was not completely clear. In our case the ghost particle creation is explicit, and rather than the steady state solution itself, of more interest to us is the approach to it. 
 
Now suppose that $\varepsilon$ is so small that the accumulated production of the two component fluid has only become significant in the recent past, so that the universe is only now approaching the steady state solution. We can determine the growth of the two fluids early in the matter dominated era when they had an insignificant effect on the evolution. To be a little more general we leave the dependence on the equation of state parameter $w_m$ of the dominant matter in the results, although we know that $w_m=0$ for this matter. We find
\begin{equation}
\rho_n={(1+w_m)t\over3+2w_n+w_m}\varepsilon,\hspace{5ex}
\rho_g=-{(1+w_m)t\over3+2w_g+w_m}\varepsilon,
\end{equation}
\begin{equation}
\rho_n+\rho_g={2(1+w_m)t\over(3+2w_n+w_m)(3+2w_g+w_m)}\varepsilon\delta w.
\label{e7}\end{equation}
The effective equation of state of this two component fluid is then
\begin{equation}
w_\mathrm{eff}={w_n\rho_n +w_g\rho_g\over \rho_n+\rho_g} = -{3\over2}-{w_m\over2}.
\end{equation}
It is interesting that this is independent of $w_n$ and $w_g$. As $\rho_n+\rho_g$ grows from insignificance in the matter dominated era and starts to affect cosmological evolution, $w_\mathrm{eff}$ will gradually increase from $-{3\over2}$. It eventually reaches $w_\mathrm{eff}=-1$ when $\rho_n+\rho_g$ comes to dominate.

To see this behavior we can numerically solve the equations and require that $\rho_n+\rho_g$ is the dark energy that is observed today, at $t=t_0=2\times10^{60}m_\mathrm{Pl}^{-1}$. This gives the result in Fig.~(1), where we see that $w_\mathrm{eff}\approx -1.2$ today. Fig.~(2) shows the behavior of $\rho_n+\rho_g$ compared with the ordinary matter density ${\rho}_m$ (with $w_m=0$). Once the product $\varepsilon\delta w$ is chosen to obtain the desired amount of dark energy, the results are quite insensitive to the individual values of $\varepsilon$ and $\delta w$. For these figures we assumed that ${\rho}_m$ and $\rho_n+\rho_g$ now make up 0.27 and 0.73 of the present total energy density, $\rho_0$, in a flat universe \cite{spergel}.
\FIGURE[t]{{\begin{picture}(300,210)
\put(0,0){\includegraphics[width=350\unitlength]{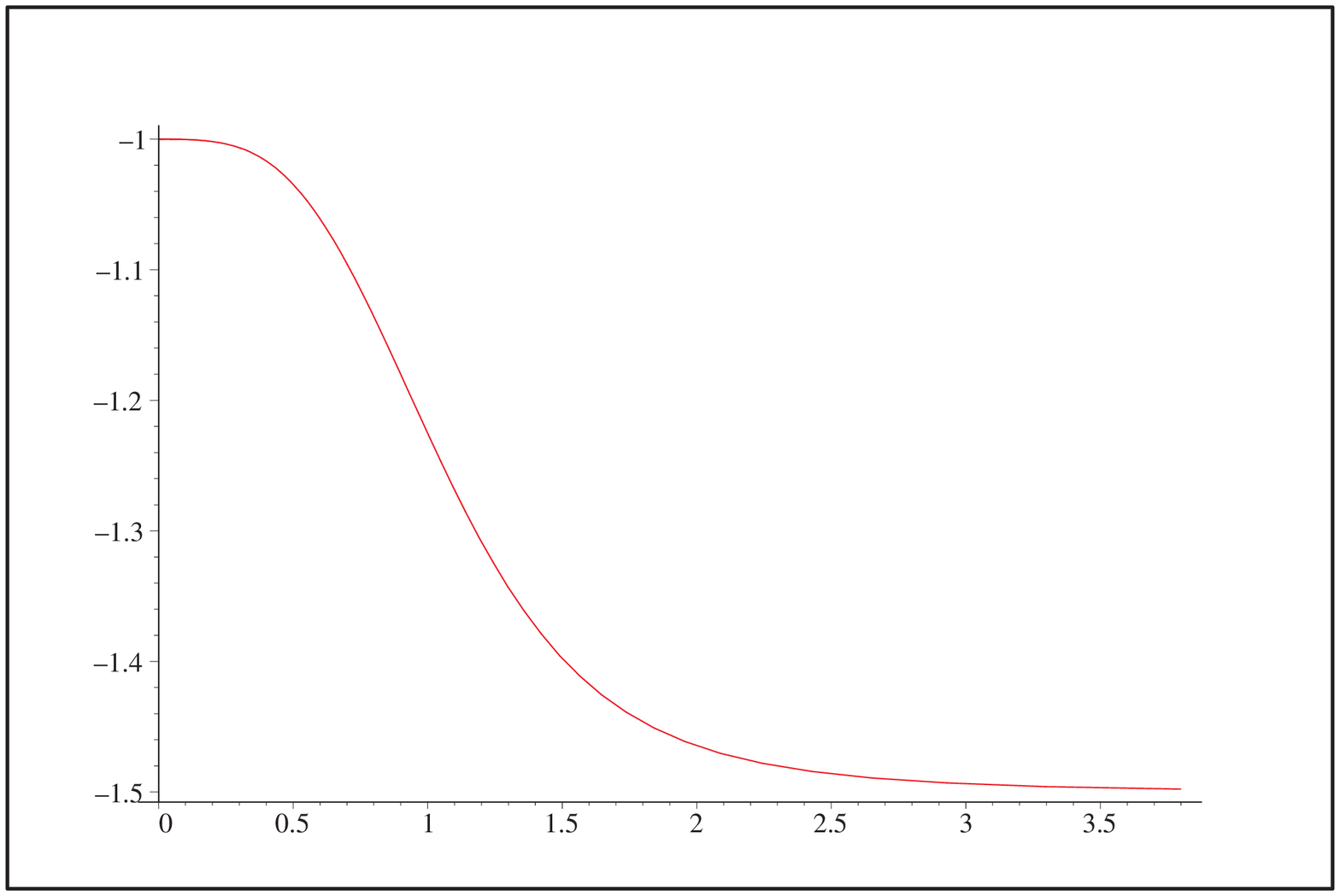}}
\put(190,7){\small$1+z$}
\put(111,23){\rule{.5\unitlength}{180\unitlength}}
\put(80,190){\small future}
\put(120,190){\small past}
\put(140,80){\Large$w_{\rm eff}$}
		  \end{picture}}%
\caption{The effective equation of state for the sum of the positive
and negative energy fluids as a function of $1+z=a(t_0)/a(t)$. The
vertical line is the present
$t=t_0=2\times10^{60}m_\mathrm{Pl}^{-1}$.\label{fig1}}}

{\renewcommand\belowcaptionskip{-1em}
\FIGURE[t]{{\begin{picture}(300,240)
\put(0,0){\includegraphics[width=350\unitlength]{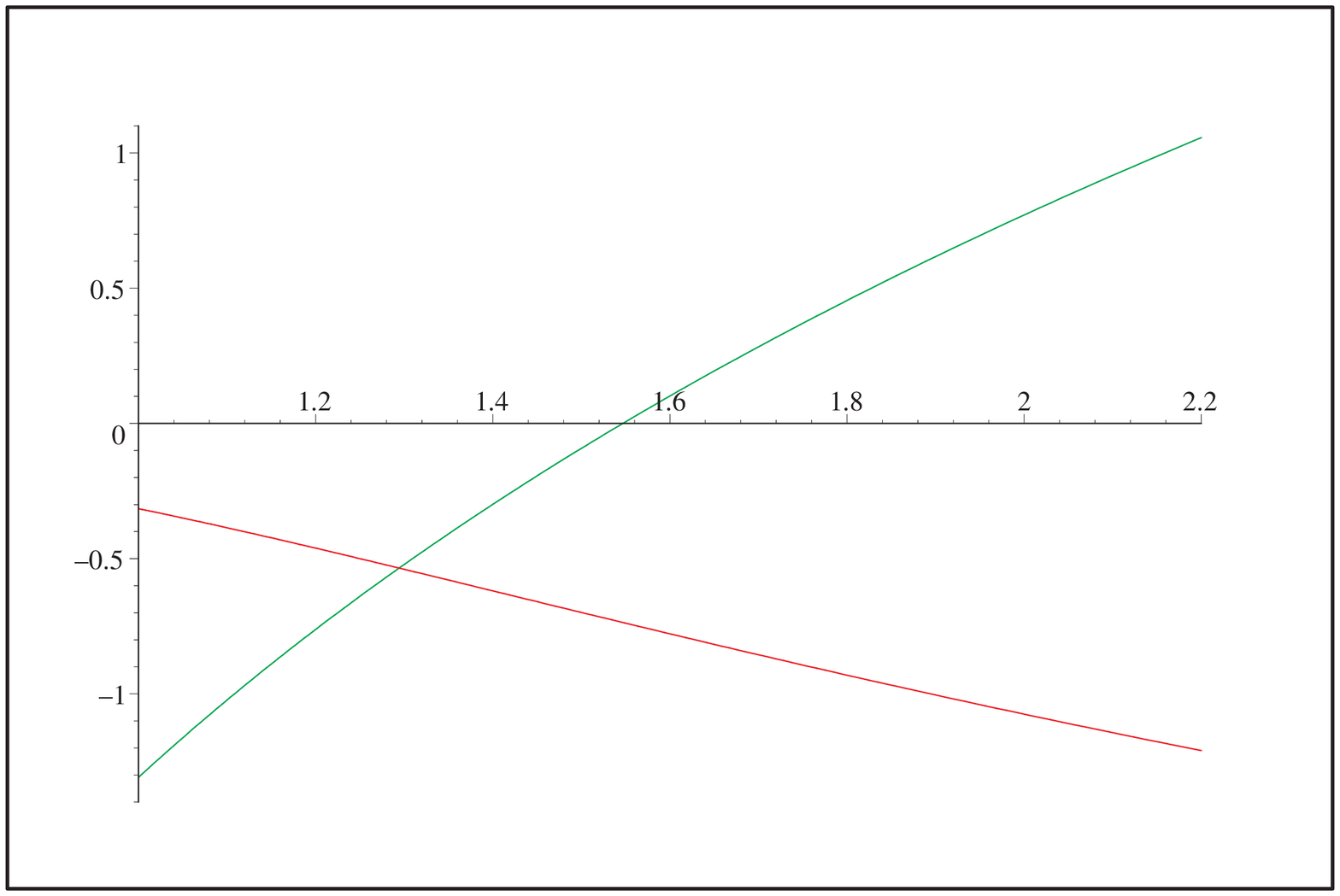}}
\put(190,110){\small$1+z$}
\put(220,180){\small $\ln(\rho_m/\rho_0)$}
\put(220,30){\small $\ln([\rho_g+\rho_n]/\rho_0)$}
		  \end{picture}}%
\caption{The energy densities compared to the present day critical
value $\rho_0$, assuming that the present day value $\rho_m/\rho_0$ is
0.27.\label{fig2}}}
}

As we have said the ghost lives in a hidden sector, and for this reason a ratio of two mass scales can produce the required suppression of $\varepsilon$. The high scale $\Lambda_H$ will be the scale of physics that couples the hidden sector to the observable sector. The low scale $\Lambda_L$ will be the ultraviolet cutoff on an effective theory in the hidden sector describing the low energy ghost degrees of freedom. It will also be the scale of Lorentz symmetry breaking, as we have anticipated in the Introduction. In the estimation of the rate of energy production per unit volume, $\varepsilon$, the phase space integrals are finite due to the cutoff $\Lambda_L$ \cite{carroll,cline}. The requirement of a Lorentz noninvariant cutoff was emphasized in \cite{cline}.

To produce the desired production rate we estimate that
\begin{equation}
\varepsilon\delta w \approx 10^{-180}m_\mathrm{Pl}^5.
\end{equation}
If $\varepsilon$ is much smaller than this then ghost production will have little impact on cosmology.\footnote{But in the next section we find that the classical evolution of the ghost field could also lead to accelerated expansion.} By dimensional arguments we have $\varepsilon \approx\Lambda_L^9/\Lambda_H^4$. For example we can first consider the case where a virtual graviton exchange provides the coupling between ghosts and nonghosts. Then $\Lambda_H$ is the Planck scale, $m_\mathrm{Pl}$, and if we ignore the $\delta w$ factor for now then ghost production can be significant today if $\varepsilon t_0\approx m_\mathrm{Pl}^2/t_0^2$. Thus $\Lambda_L^3\approx m_\mathrm{Pl}^2/t_0$, which implies that $\Lambda_L$ is of order 10 MeV or so. We will be led to consider different values of $\Lambda_H$ and $\Lambda_L$ in section 4.

As for $\delta w$, we note that some of the normal particles produced can be massive, and so even if the normal particles are relativistic the total effective $w_n$ can be slightly less than $1/3$. If the ghost was massless and had a standard relativistic equation of state, $w_g=1/3$, then we would estimate that $\delta w\approx m_h^2/\Lambda_L^2$, where $m_h$ is the most massive of the normal particles. Thus a massless ghost is one way to ensure that $\delta w>0$. Given this it is becoming clear that we should develop the following picture: at some dynamical scale $\Lambda_L$, Lorentz invariance is spontaneously broken in a hidden sector, and a massless Goldstone boson emerges as a ghost. We will explore a natural connection between Lorentz symmetry breaking and a Goldstone ghost in the next section. One consequence of this picture is that $w_g$ is likely larger than $1/3$.

Ghost scalar fields (phantoms) have been proposed before to account for a possible $w<-1$ equation of state \cite{caldwell}. In that picture a ghost field is presently rolling on a potential, with the requirement that the ghost mass is extremely small but nonzero, of order $H_0\approx10^{-33}$ eV. In our picture this type of mass scale is not necessary and the ghost can be massless.

\section{The Goldstone Ghost}

Let us explore the connection between ghosts and Lorentz symmetry breaking in a phenomenological model of the physics below the $\Lambda_L$ scale. We can follow the discussion of the ``ghost condensate'' in \cite{nima}, but with a crucial change of sign. We start with a Lagrangian of a scalar with a shift symmetry and with an arbitrary numbers of derivatives of the form
\begin{equation}
{\cal L}=\Lambda_L^4 P(X)\hspace{3ex}\mbox{where}\hspace{3ex}X={\partial^\mu\phi\partial_\mu\phi\over\Lambda_L^4}
.\label {e8}\end{equation}
But unlike \cite{nima} we choose $P'(0)>0$, making this a scalar field theory with the \textit{right} sign kinetic term, ${\cal L}={1\over2}\partial^\mu\phi\partial_\mu\phi+...$ . As in \cite{nima} we consider small Goldstone fluctuations around Lorentz violating solutions,
\begin{equation}
\phi(x)=c\Lambda_L^2\,t+\pi(x).
\end{equation}
We assume that there is a positive value of $c^2$ where $P'(c_*^2)=0$ and $P''(c_*^2)<0$, in which case we have
\begin{equation}
{\cal L}=2c_*^2P''(c_*^2)\,\dot{\pi}^2.
\end{equation}
Since $P''(c_*^2)<0$ this Lagrangian describes a ghost excitation. There is no $(\nabla \pi)^2$ term, as in \cite{nima}. Now by considering more general terms in (\ref{e8}), again with the opposite signs to what was chosen in \cite{nima}, we can end up after a rescaling with a Lagrangian for the Goldstone ghost of the form
\begin{equation}
{\cal L}=-{1\over2}{\dot\pi}^2+{1\over2}{\bar{\Lambda}_L^2\over\Lambda_L^4}(\nabla^2 \pi)^2,
\label{e9}\end{equation}
with $\bar{\Lambda}_L\approx\Lambda_L$. This is the same Lagrangian considered in \cite{nima}, except for the opposite overall sign.  If we turn off gravity by taking the limit $m_\mathrm{Pl}\rightarrow\infty$ we are left with a sensible theory, just as in \cite{nima}, since then the overall sign is of no consequence. All the implications that follow in this limit will be the same. The different sign will only have consequences when the coupling to gravity or anything else is considered, and in particular it leads to the instability associated with ghost production from the vacuum that we are considering in this paper.

Although the different overall sign in (\ref{e9}) was chosen in \cite{nima} to avoid this instability, nevertheless a different instability was discovered in the modes of $\pi$ with some characteristic length scale, due to the coupling to gravity. This is reflected by a negative term in the dispersion relation, arising when the mixing between $\pi$ and the scalar perturbation of the metric is taken into account.
\begin{equation}
\omega^2={\bar{\Lambda}_L^2\over\Lambda_L^2}\left({k^2\over\Lambda_L^2}-{\Lambda_L^2\over2 m_\mathrm{Pl}^2}\right) k^2.
\end{equation}
Requiring that the time scale of this instability is greater than the age of the universe implies that $\Lambda_L^3\lesssim m_\mathrm{Pl}^2/t_0\approx\mbox{(10 MeV)}^3$. Perhaps not surprisingly, this is the same constraint we have found for the ghost production instability. Meanwhile, with our choice of sign in (\ref{e9}) the dispersion relation becomes
\begin{equation}
\omega^2={\bar{\Lambda}_L^2\over\Lambda_L^2}\left({k^2\over\Lambda_L^2}+{\Lambda_L^2\over2 m_\mathrm{Pl}^2}\right) k^2.
\end{equation}
Thus the instability in the dispersion relation is removed in our picture, and the phenomena associated with this instability, such as exponentially growing oscillations in Newtonian potentials, no longer appear.

We can follow the calculation in \cite{nima} of the correction to the Newtonian potential that develops after a time $t$, but with the different sign incorporated. We find
\begin{equation}
\Phi(r,t)=-{G\over r}[1+I(r,t)],
\end{equation}
\begin{equation}
I(r,t)\approx-.364\,T\,g(R^2/T),
\end{equation}
for small $T$ where
\begin{equation}
T={\bar{\Lambda}_L^2\Lambda_L\over 2m_\mathrm{Pl}^2}t,\hspace{5ex}R={\Lambda_L^2\over \sqrt{2}m_\mathrm{Pl}}r.
\end{equation}
The maximum value of $|g|$ is at $g(1)=1$. Thus, although there is no instability here, there is some upper bound on $T$ so that these corrections are not too large. This again requires that $\Lambda_L^3$ is sufficiently small compared to $m_\mathrm{Pl}^2/t_0$. We do not pursue these types of effects here because of the limit on $\Lambda_L$ that will emerge in the next section.

It is of interest to consider a nonvanishing $P'(X)$, in particular when $X=\dot{\phi}^2/\Lambda_L^4>c_*^2$. For $X$ not too much larger than $c_*^2$ then the Goldstone excitation is still ghost-like, and $P'(X)<0$, since $P''(c_*^2)<0$. In this situation there is a new $-P'(X)(\nabla\pi)^2$ term in the ghost Lagrangian which is of the correct sign to ensure classical stability within the ghost sector. (This would not have been the case for $X<c_*^2$.) The classical ghost contributions to the energy-momentum tensor are then
\begin{equation}
\rho=\Lambda_L^4(2XP'(X)-P(X)),\;\;\;\;\;p=\Lambda_L^4P(X).
\end{equation}
If we further assume that $P(X)<0$ then we have here a positive cosmological constant contribution plus a dust-like negative energy contribution. The result is that the equation of state parameter for this Goldstone ghost condensate is $w\equiv\rho/p<-1$. The dust-like component will dilute away under cosmological expansion, and thus $w\rightarrow-1$ at later times.

This gives an alternative picture for $w<-1$ dark energy, possibly of interest if the ghost production mechanism was too slow. In this picture, to have $w$ appreciably smaller than $-1$ at present would seem to require a fairly recent breaking of Lorentz symmetry breaking. In other words $\Lambda_L^4$ should be very small, of order the observed dark energy density, and this would make it quite unlikely to see any other effect of the Goldstone ghost sector. In the following we do not consider further the implications of a nonvanishing $P'(X)$,\footnote{We note though that this negative energy substance is accreted by black holes, just as the positive energy ghost condensate is \cite{frolov}. It affects the largest black holes first, and it causes them to shrink instead of grow.} and the total cosmological constant is assumed to vanish.

\section{Ghost production}

We have seen how a theory with a good perturbative vacuum can become trapped in a region of field space which breaks Lorentz symmetry and yields a Goldstone ghost. Because of the coupling to the hidden sector at scale $\Lambda_H$, ghosts and normal particles are produced at a constant rate from the vacuum with energies below $\Lambda_L$. The ghosts can be pictured as fluctuating disturbances of the constant $\dot{\phi}$ solution, with the highest frequencies possible being set by $\Lambda_L$. This Lorentz violating statement is related to the existence of a preferred frame. In fact the approximation to the dispersion relation, $\omega^2\approx k^4/\Lambda_L^2+...$, must be breaking down as the group velocity $v^2\approx k^2/\Lambda_L^2$ approaches the speed of light.

In any case the redshifting of the wave vectors $k\rightarrow k/a(t)$ from expansion implies a redshifting of the energies that can be quite different than $a(t)^{-1}$, and in fact is likely faster. For example for $\omega$ such that $\omega^2=k^4/\Lambda_L^2$ is a good approximation, the energy \textit{density} in the ghost modes should redshift as $a(t)^{-2}a(t)^{-3}$. This corresponds to $w_g=2/3$ and $\delta w\approx1/3$. We will assume this in the following even though the behavior of the highest frequency modes may be somewhat different. In fact any $\delta w>0$ is acceptable.

We can consider the rate of production of ghosts and normal particles occurring today, since it is here that observational bounds come into play. The strongest bound comes from photons \cite{cline}. The production implies a differential flux density for photons today that peaks as a function of $E$ at
\begin{equation}
{dF\over dE}\approx {\varepsilon f_\gamma t_0\over\Lambda_L^2}\hspace{3ex}\mathrm{for}\hspace{3ex}E\approx\Lambda_L.
\end{equation}
$f_\gamma$ is the photon branching fraction in the total production. The measured flux density is approximately of the form ${C/E^2}$ \cite{sreek}, and so this results in a simple constraint on $\varepsilon f_\gamma$,
\begin{equation}
\varepsilon f_\gamma\lesssim C/t_0\approx 10^{-191}m_\mathrm{Pl}^5.
\end{equation}
On the other hand we have seen that to get sufficient dark energy $\varepsilon \approx 10^{-180}m_\mathrm{Pl}^5$, which in turn would require that $f_\gamma\lesssim10^{-11}$. This is unreasonable because of the universal couplings of gravity, and so this basically rules out the picture where gravity is responsible for the coupling of ghosts to normal matter.

We will thus consider lowering $\Lambda_L$ so that some other interaction can be more important than gravity for the ghost production. $\Lambda_L$ is then well below the electron threshold, leaving photons and neutrinos to be produced along with the ghosts. We will discuss how it is that the branching fraction of photons can be suppressed. Consider a broken gauge interaction with a mass scale $\Lambda_H$ which couples the hidden sector to known particles. Unlike the graviton exchange, here the physics responsible for the coupling is massive and is integrated out at the $\Lambda_H$ scale, leaving us to consider the induced operators in the low energy theory. A few of interest involve leptons and the photon,
\begin{equation}
F^\mu(\Psi_\mathrm{hidden})\left[\overline{\nu}_L\gamma_\mu\nu_L+\overline{e}\gamma_\mu (a+b\gamma_5) e\right],
\label{e1}\end{equation}
\begin{equation}
G(\Psi_\mathrm{hidden})F^{\mu\nu}F_{\mu\nu}.
\end{equation}
$\Lambda_H$ should still be above the electroweak scale, and thus $a-b=1$ by SU$_L$(2) symmetry. $F^\mu$ and $G$ contain fields from the hidden sector such that when the operators are run down to a scale below $\Lambda_L$, they induce interactions involving ghost fields. $F^\mu$ and $G$ also contain appropriate powers of $1/\Lambda_H$. If they both involve two fermions then the lepton operator is suppressed by $1/\Lambda_H^2$, while the photon operator will have $1/\Lambda_H^3$ or even more suppression. Thus the contribution from the $G$ term to the photon branching fraction is suppressed by at least $\Lambda_L^2/\Lambda_H^2$. This will turn out to be negligible.

We also have to consider the electron term of the operator (\ref{e1}). This will induce a two photon term when it is run down to the scale $\Lambda_L$, which is now far below the electron mass. This will arise from an electron triangle graph which will be proportional to the $b$ term in (\ref{e1}). The new photon operator must involve $\varepsilon_{\mu\nu\rho\sigma}$, two factors of $F_{\mu\nu}$ by gauge invariance, and an extra derivative by the Lorentz invariance of the observable sector. This implies a factor of $1/m_e^2$ in the amplitude, and thus a photon branching fraction $f_\gamma$ of order $\alpha^2\Lambda_L^4/m_e^4$. Combining this with the observational constraint $f_\gamma\lesssim10^{-11}$, along with the fact that neutrinos need to be produced, we obtain
\begin{equation}
\mbox{(smallest neutrino mass)}\lesssim\Lambda_L\lesssim10\,\mathrm{KeV}.
\label{e4}\end{equation}

On the other hand $\Lambda_H$ and $\Lambda_L$ are constrained in order to obtain the right amount of dark energy, $\varepsilon\approx\Lambda_L^9/ \Lambda_H^4 \approx 10^{-180}m_\mathrm{Pl}^5$. An interesting value would be $\Lambda_H\approx 1000$ TeV, since this is the natural scale for flavor physics in dynamical theories of quark and lepton masses, and thus could be the natural scale for coupling known fermions, neutrinos in particular, to an unobserved sector. This would then imply that $\Lambda_L\approx100$ eV.

We can also consider the explicit Lorentz violating effects that feed into the observable sector through the interactions at scale $\Lambda_H$. The leading effects are again from operators as in (\ref{e1}) with coefficients of order $1/ \Lambda_H^2$. Then the Lorentz violating terms, an example of which is the electron term $\overline{e}\gamma_\mu\gamma_5e$, will have coefficients of order $\Lambda_L^3/ \Lambda_H^2$. For $\Lambda_H\approx 1000$ TeV and $\Lambda_L\approx100$ eV, the result is about 6 orders of magnitude smaller than the current bound \cite{muller}. We conclude that there is quite a wide range for the scales $\Lambda_L$ and $\Lambda_H$ that produce desirable results.

\section{Conclusion}

A consistent infrared modification to gravity involving Lorentz symmetry breaking results in some kind of instability. Very schematically, the mixing between the Goldstone boson $\pi(x)$ and the nonpropagating scalar mode of gravity $\Phi(x)$ is given by \cite{nima}
\begin{equation}
{\cal L}=-{1\over2}(\nabla\Phi)^2\pm{1\over2}(m\Phi-\dot{\pi})^2+...
\end{equation}
with $m=\Lambda_L^2/\sqrt{2}m_\mathrm{Pl}$. The choice of the positive sign in this Lagrangian ensures that the Goldstone boson has positive kinetic energy, but it gives the nonstandard sign for the mass term of $\Phi$. This nonstandard sign and the ensuing mixing leads to a classical instability, yielding the exponential growth of modes characterized by a length scale $m_\mathrm{Pl}/\Lambda_L^2$ and a time scale $m_\mathrm{Pl}^2/\Lambda_L^3$. In this paper we have instead chosen the minus sign and thus a Goldstone ghost, which does not suffer from this particular instability. There are still gravitational implications of the mixing, and they are still characterized by the same length and time scales as before, but they are now well behaved.\footnote{At the end of section 3 we also briefly noted that the classical Goldstone ghost field can directly yield a $w<-1$ dark energy component that evolves towards $w=-1$ in the future.}

The Goldstone ghost gives rise to a different instability, that of ghost production from the vacuum. This has a finite rate because of the cutoff provided by the scale of Lorentz symmetry breaking. If it is graviton exchange that is responsible for the instability then the time scale for its effects to show up is again $m_\mathrm{Pl}^2/\Lambda_L^3$. It appears that the instability of the Goldstone ghost is of particular cosmological interest, since it pushes the scale factor of the whole universe towards exponential growth. The constant particle production from the vacuum balances against the dilution from the exponential growth, and thus the endpoint of this ``instability'' is well behaved. This picture is quite independent of the remaining question of why the cosmological constant vanishes.

If ghost production is the cause of the present accelerated expansion, then the universe is only part way in its transition to the steady state behavior. The effective equation of state parameter of the produced ghosts and normal particles evolves from $-3/2$ to $-1$ and is presently $w_\mathrm{eff}\approx -1.2$. The ghosts are basically invisible, but the normal particles produced lead to observational constraints. In fact the normal particles should be predominantly neutrinos rather than photons, and for this reason we considered a nongravitational coupling between the hidden sector and our own. In the case of a broken gauge interaction, perhaps connected to flavor physics around 1000 TeV, we showed that adequate suppression of photons relative to neutrinos can naturally occur. The energy spectrum of neutrinos will peak around the scale of Lorentz symmetry breaking, $\Lambda_L$, which must be below 10 KeV.

$\Lambda_L$ sets an upper bound on the energy that a propagating ghost can carry. Above this scale there is some ultraviolet completion of the infinite derivative scalar field model. We speculate that this more fundamental theory in the hidden sector could contain stable massive states with positive energy. This would have to be consistent with having a sufficiently stable ground state in the hidden sector (when considered in isolation). This may be possible as long as there is sufficient separation between the minimum mass of these states and the maximum energy (in absolute value) that can be carried by the Goldstone ghosts. Positive energy in the hidden sector may then end up in these massive states as the universe cools, and we note that the mass and number density of these states may make them of interest for dark matter. 

Finally we note that the proximity of $\Lambda_L$ to the scale of neutrino masses may suggest a common origin for these mass scales, perhaps related to the physics of flavor.

\acknowledgments
I thank T. Hirayama for numerous discussions on this and related work. This research was supported in part by the Natural Sciences and Engineering Research Council of Canada.
\appendix
\section{Massive Ghosts}

In this appendix we briefly consider classical ghost dynamics involving scalar field theories with arbitrarily large masses. Although we are again looking for exponential expansion, this discussion is quite unrelated to the rest of this paper. Of interest is the way that gravity can induce a coupling between the ghost and normal field at the classical level, and it is also curious to see the role that a cosmological constant plays in these solutions.

Consider two free fields, a ghost scalar $\phi_{g}$ and a normal scalar $\phi_{n}$,
\begin{equation}
{\cal L}={1\over2}\partial_\mu \phi_n\partial^\mu \phi_n-{1\over2}m_n^2\phi_n^2
-{1\over2}\partial_\mu \phi_g\partial^\mu \phi_g+{1\over2}m_g^2\phi_g^2
,\end{equation}
along with the scale factor $a(t)$ in a spatially flat metric. The only coupling between the two fields is provided through their coupling to gravity; the two fields can only interact if $a(t)$ is nonconstant. For solutions of the form $\phi_{n}=\sigma(t)\cos(\mathbf{k}_{n}\cdot\mathbf{x})$, $\phi_{g}=\eta(t)\cos(\mathbf{k}_{g}\cdot\mathbf{x})$, and after performing a spatial average of the ghost and scalar contributions to $T_{\mu\nu}$, the two field equations along with the Einstein equation gives three equations for the three quantities $a(t)$, $\sigma(t)$, $\eta(t)$. The effects of a cosmological constant $\Lambda$ is included.

We first consider the case of $m_g=m_n$ and $k_g=k_n$. The oscillating fields induce oscillations in $a(t)$ which in turn induces a coupling between the fields. The size of this coupling is proportional to the amplitudes and thus in this case we find exponentially growing amplitudes for $\phi_n$ and $\phi_g$. The time dependent $a(t)$ that supports this behavior can be oscillatory and well-behaved. For this to be the case it turns out that $\Lambda$ must be negative. In fact $\sqrt{|\Lambda|}$ plays the role of a coupling between the positive and negative energy modes, since it sets the rate of exponential growth of the field amplitudes.

More interesting is the unequal mass case where $m_n>m_g$. (This is analogous to the $w_g>w_n$ constraint discussed in this paper.) Here the ghost and scalar modes with arbitrary choices of $k_g$ and $k_n$ typically do not grow in an uncontrolled way. Instead the amplitudes of these oscillations are found to be of order $\sqrt{|\Lambda|}/m_n$, where $\Lambda$ is again negative. It is when this ratio is not much smaller than unity that we find solutions where the scale factor experiences exponential growth. de Sitter expansion occurs even though the cosmological constant is negative. This occurs because of a flow of energy between the positive and negative energy modes, due to their common coupling to gravity, so that their respective amplitudes remain fixed while expansion occurs. Fine-tuning is not required for these solutions, and the parameters in the theory determine the expansion rate.

\end{document}